\documentclass[conference,letterpaper]{IEEEtran}
\usepackage{bbding}
\usepackage[T1]{fontenc}
\usepackage{aurical}
\usepackage{huncial}
\usepackage{linearb}
\usepackage{textcomp}
\usepackage{amssymb,epic,eepic}
\usepackage{amsmath}
\usepackage[amsmath,amsthm,thmmarks]{ntheorem}
\usepackage{amsfonts}
\usepackage{stmaryrd}
\usepackage{capt-of}
\usepackage{graphicx}
\usepackage[usenames,dvipsnames]{pstricks}
\usepackage{epsfig}
\usepackage{pst-grad}
\usepackage{pst-plot}
\usepackage{mathtools}
\usepackage{textfit} 
\usepackage{subfigure}
\usepackage{float}

\usepackage{comment}

\newcommand{\PP}{{\mathbb{P}}}

\newcommand{\n}{{\mathbf n}}

\newcommand{\cB}{\mathcal{B}}
\newcommand{\cC}{\mathcal{C}}
\newcommand{\cE}{\mathcal{E}}

\newcommand{\cN}{\mathcal{N}}
\newcommand{\cM}{\mathcal{M}}
\newcommand{\cS}{\mathcal{S}}
\newcommand{\cT}{\mathcal{T}}
\newcommand{\cU}{\mathcal{U}}
\newcommand{\cX}{\mathcal{X}}

\newcommand{\cZ}{\mathcal{Z}}

\newcommand{\Tr}{\operatorname{Tr}}

\newcommand{\id}{\operatorname{id}}

\newcommand{\ox}{\otimes}
\newcommand{\1}{\mathbf{1}}

\newcommand{\ket}[1]{\left|#1\right\rangle}
\newcommand{\bra}[1]{\left\langle#1\right|}
\newcommand{\proj}[1]{\left|#1\right\rangle\!\left\langle#1\right|}

\newtheorem{theorem}{Theorem} 

\newtheorem{definition}[theorem]{Definition}
\newtheorem{remark}[theorem]{Remark}

\ifCLASSINFOpdf
\else
\fi
\begin{document}

\title{Quantum Wiretap Channel Coding \protect\\ 
Assisted by Noisy Correlation}

\author{\IEEEauthorblockN{Minglai Cai\textsuperscript{1}}
\IEEEauthorblockA{
\textsuperscript{1}\textit{Grup d'Informaci\'o Qu\`antica}\\
\textit{Departament de F\'isica}\\
\textit{Universitat Aut\`onoma}\\
\textit{de Barcelona, Spain}\\
minglai{\_}cai@hotmail.com} 

\and
\IEEEauthorblockN{ }
\IEEEauthorblockA{
\textsuperscript{2}\textit{ICREA--Instituci\'{o}}\\
\textit{Catalana de Recerca}\\
\textit{i Estudis Avan\c{c}ats}\\
\textit{Barcelona, Spain}}

\and
\IEEEauthorblockN{ }
\IEEEauthorblockA{
\textsuperscript{3}\textit{Institute for}\\
\textit{Advanced Study}\\
\textit{Technische Universit\"at}\\
\textit{M\"unchen, Germany}} 

\and
\IEEEauthorblockN{Andreas Winter\textsuperscript{1,2,3,4}}
\IEEEauthorblockA{
\textsuperscript{4}\textit{QUIRCK--Quantum}\\
\textit{Independent Research}\\ 
\textit{Center Kessenich}\\
\textit{Bonn, Germany}\\
andreas.winter@uab.cat}
}

\maketitle


\begin{abstract}
We consider the private classical capacity of a quantum wiretap 
channel, where the users (sender Alice, receiver Bob, and eavesdropper Eve) 
have access to the resource of a shared quantum state, additionally
to their channel inputs and outputs. 
An extreme case is maximal entanglement or a secret key between
Alice and Bob, both of which would allow for one-time padding the message. 
But here both the 
wiretap channel and the shared state are general. In the other
extreme case that the state is trivial, we recover the 
wiretap channel and its private capacity 
[N. Cai, A. Winter and R. W. Yeung, \emph{Probl. Inform. Transm.} 
40(4):318-336, 2004]. 
We show how to use the given resource state
to build a code for secret classical communication. 
Our main result is a lower bound on the assisted private capacity, 
which asymptotically meets the multi-letter converse and which encompasses 
all sorts of previous results as special cases.
\end{abstract}


\section{Introduction}
Entanglement shared between sender and receiver of a transmission is a
useful resource that generically increases channel capacity.
In this scenario, the two parties may be given some number of quantum 
bits jointly prepared in a fixed superposition.
The advantages of a shared quantum entanglement 
resource prior to communication over  a noisy quantum channel has been
considered extensively in \cite{Be/Wi} and \cite{Be/Sh/Sm/Th,Be/Sh/Sm/Th2}, 
where the entanglement-assisted classical capacity theorem
has been derived; see also \cite{Wa/Ha}. 
These works showed how to increase the classical capacity of quantum channels
by the assistance of unlimited pure entanglement. 
Since then, more recent works have studied extensions of the use of 
shared  entanglement as an assisting quantum resource:
entanglement-assisted communication over quantum multiple-access channels
\cite{Hs/De/Wi};
entanglement-assisted communication over compound and 
arbitrarily varying quantum channels \cite{Bo/Ja/Ka};
entanglement-assisted communication over quantum broadcast channels
\cite{Qi/Sh/Wi}.
For these results, it is essential that
the shared state be  maximally entangled.
In a somewhat dual setting, 
in \cite{Bo/Pl/Ve,Bow,Hiro,Ho/Ho/Ho/Le/Te,Win2001,Ho/Pi}, 
Alice and Bob are connected by a noiseless quantum channel, but instead share a general 
mixed quantum state (decoupled from Eve). 
The case of a noisy quantum channel has been investigated in \cite{Ba/Wi/Ya}.
These papers showed that sufficiently entangled states, or any entanglement in conjunction with suitable channels, are useful resources for communication.

Secure communication over a classical channel with an eavesdropper was first introduced by Wyner \cite{Wyn}.
The secrecy capacity for quantum wiretap channels 
has been determined in \cite{De} and \cite{Ca/Wi/Ye}.    
%
Chen \emph{et al.} \cite{Ch/Ca/Se} demonstrated that correlation between 
Alice and Bob is able to help secure message transmission
over a classical wiretap channel.
Inspired by these results, we analyze secure communication 
when the channel users (Alice, Bob and Eve) share a general correlated state,
which is ``given by nature'', and assumed to be known to all parties.
In this work, we show a way to increase 
the secure capacity of quantum channels
by the assistance of such noisy correlation. 
Our proof works by considering the shared resource as
a component of the channel.
Technically, we employ quantum wiretap channel codes, and incorporate the correlated resource via a variant of Gel'fand-Pinsker coding as in \cite{Ch/Vi} and \cite{An/Ha/Wa}. This has been used similarly in secure 
``writing on dirty paper'' codes \cite{Co}.

\begin{figure}[ht]
\begin{center}
\includegraphics[width=0.99\linewidth]{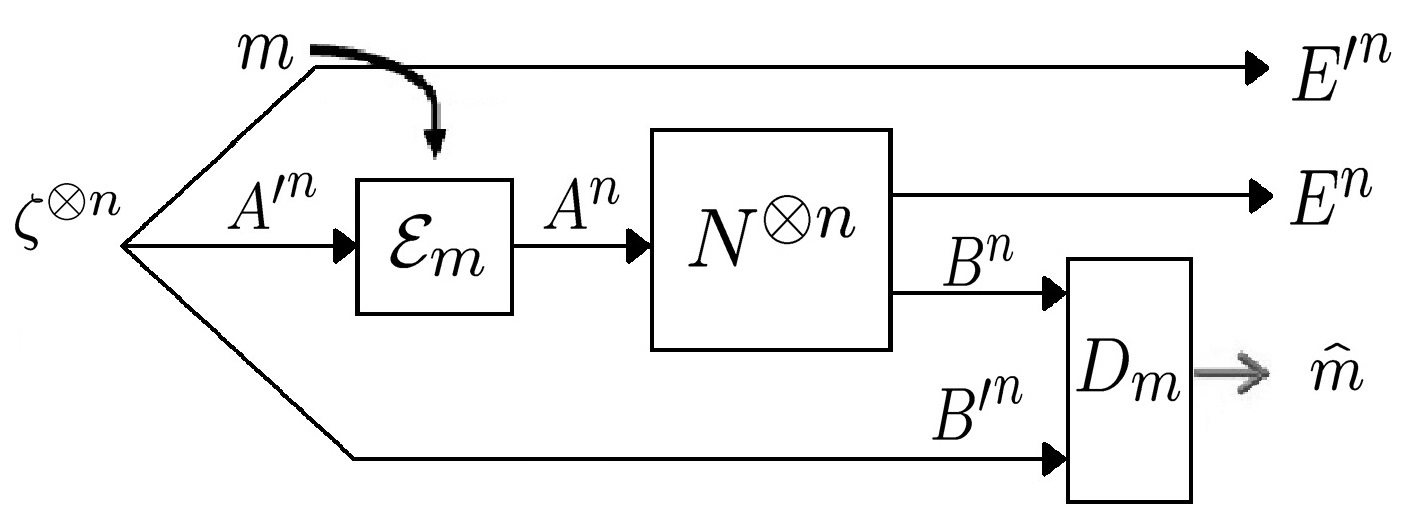}
\caption{Communication diagramme of the wiretap coding problem over $\cN$ with assistance by a tripartite resource state shared by sender (Alice), receiver (Bob) and eavesdropper (Eve). Unlike the plain wiretap channel, here each message is encoded not into a state, but rather a modulation $\mathcal{E}_m:A'\rightarrow A$.}
\label{wireqresouima2}
\end{center}
\end{figure}

We would like to mention another coding protocol, which has appeared previously in the literature, showing a similar result \cite{Wil/Hsi}: this paper determined the entanglement generation capacity of a quantum channel with shared quantum states as a resource. The proof of the achievability part uses a similar approach to the dirty paper codes, although our construction cannot be reduced to theirs, in the same way as secret key generation is more general than entanglement generation \cite{HHHO-key}. 
See also \cite{De} for the first description of codes achieving the entanglement generation capacity based on wiretap codes. 

In the next section we introduce the communication scenario and the assisted private capacity, followed by a section recalling certain basic notions. After that we state and prove our main result, followed by a discussion of special cases of interest.

\section{Communication scenario}
\label{scene}
Before starting, we review some basic notions from quantum information
theory.
Let $\rho_1$ and $\rho_2$ be Hermitian operators on a  finite-dimensional
complex Hilbert space $A$. We say $\rho_1\geq\rho_2$ and $\rho_2\leq\rho_1$ 
if $\rho_1-\rho_2$ is positive semidefinite.
We denote the set of  density operators (states) on $A$ by
$\cS(A) := \{\rho \in \mathcal{L}(A) : \rho \geq 0,\  \Tr(\rho) = 1 \}$,
where $\mathcal{L}({A})$ is the set  of linear  operators on ${A}$. 
For finite-dimensional complex Hilbert spaces $A$ and $B$,  
a \emph{quantum channel} is a linear, 
completely positive and trace preserving (cptp) map 
$\cN:\mathcal{L}(A) \rightarrow \mathcal{L}(B)$, that acts a 
$\mathcal{L}(A) \ni \rho \mapsto \cN(\rho) \in \mathcal{L}(B)$, 
which accepts input quantum states in $\mathcal{S}(A)$ and produces 
output quantum states in $\mathcal{S}(B)$. We shall often simplify 
the notation of a quantum channel as $\cN:A\rightarrow B$. Likewise, 
we will often suppress the tensor product sign in $BE = B\ox E$
and $A^n=A^{\ox n}$, if there is no danger of confusion.

\begin{definition}
\label{def:q-wiretap-channel}
A \emph{quantum wiretap channel} is a cptp map $\cN: A \rightarrow B\otimes E$, 
with finite-dimensional quantum systems $A$, $B$ and $E$, representing Alice's input and Bob's and Eve's outputs, respectively. 
\end{definition}
Note that in \cite{Ca/Wi/Ye}, a wiretap channel was more generally a pair 
of cptp maps, one from $A$ to $B$ and one from $A$ to $E$. We can reproduce that notion by letting $\cN_B=\Tr_E\circ\cN$ and $\cN_E=\Tr_B\circ\cN$, but not every pair of cptp maps arises in this way. However, the generality offered by \cite{Ca/Wi/Ye} comes in handy later on. 

\begin{definition}
\label{def:q-wiretap+zeta}
A \emph{quantum wiretap channel assisted by a correlation} is a pair 
$(\cN,\zeta)$ consisting of a wiretap channel
$\cN:A \rightarrow B\otimes E$ and a 
quantum state $\zeta \in \cS(A'\ox B'\ox E')$, with finite-dimensional 
quantum systems $A'$, $B'$ and $E'$ in the possession of Alice, Bob and 
Eve, respectively, before the transmission. 
\end{definition}

\begin{definition}
\label{def:code}
An \emph{$(n,\lambda,\mu)$-wiretap code} for $(\cN,\zeta)$ is a collection 
$\{(\cE_m,D_m) : m\in[M]=\{1,\ldots,M\}\}$ of cptp maps
$\cE_m:{A'}^n\rightarrow A^n$ (the ``modulations'') and operators 
$D_m \geq 0$ on $B^n{B'}^n$, $\sum_{m=1}^M D_m \leq \1$, such that 
there exists a state $\sigma$ on $E^n{E'}^n$ with
\begin{align}
  \label{eq:reliable}
  \frac1M &\sum_m \Tr\left( (\cN^{\ox n}\!\circ\cE_m)\zeta^{\ox n}\!\cdot\!(D_m\!\ox\1_{E^n{E'}^n})\right) \geq 1-\lambda, \\
  \label{eq:private}
  \frac1M &\sum_m \left\| \Tr_{B^n{B'}^n}(\cN^{\ox n}\!\circ\cE_m)\zeta^{\ox n} - \sigma^{E^n{E'}^n} \right\|_1 \leq \mu.
\end{align}
See Fig. \ref{wireqresouima2} for the communication diagram. 
\end{definition}

The \emph{rate} of the wiretap code is $\frac1n\log M$, and the largest number 
$M$ of 
messages of an $(n,\lambda,\mu)$-wiretap code is denoted $M(n,\lambda,\mu)$. 
This allows us to define the (weak) \emph{assisted private capacity} as
\begin{equation}
\label{eq:assisted-private}
P(\cN,\zeta) := \inf_{\lambda,\mu>0} \liminf_{n\rightarrow\infty} \frac1n\log M(n,\lambda,\mu), 
\end{equation}
which is the main objective for the rest of the paper.

\section{Preliminaries}
\label{prel}
For a finite set $\cX$, we denote the
set of probability distributions on $\mathcal{X}$ by $\PP(\mathcal{X})$.
For a discrete random variable $X$  on a finite set $\mathcal{X}$ and a discrete
random variable  $Y$  on  a finite set $\mathcal{Y}$  we denote the Shannon entropy
of $X$ by
$H(X)=-\sum_{x \in \mathcal{X}}p(x)\log p(x)$ and the mutual information between $X$
and $Y$ by  
\[I(X;Y) = \sum_{x \in \mathcal{X}}\sum_{y \in \mathcal{Y}}  p(x,y) \log{\frac{p(x,y)}{p(x)p(y)}}\text{ ,}\]
Here $p(x,y)$ is the joint probability distribution function of $X$ and $Y$, and 
$p(x)$ and $p(y)$ are the marginal probability distribution functions of $X$ and $Y$ respectively.
Throughout the paper the logarithm base   is  $2$.

For quantum states acting on $A$ or composite systems $A \ox B$, these concepts generalise. 
For instance for $\rho^{AB} \in \cS(A\ox B)$ we have the marginals (reduced states) $\rho^A = \Tr_B\rho$ and $\rho^B=\Tr_a\rho$ given by the partial trace. To a finite set $\cX$, we associate the Hilbert space $X$ with orthonormal basis $\{\ket{x}:x\in\cX\}$, such that the possible distributions $p(x)$ on $\cX$ correspond to diagonal density matrices $\rho = \sum_x p(x) \proj{x}$. 

An ensemble $\{p(x),\rho_x\in\cS(A)\}_{x\in\cX}$ of states on $A$ is faithfully represented by the \emph{classical-quantum (cq-)state} 
\[
  \gamma = \sum_{x\in\cX} p(x) \proj{x}^X \ox \rho_x^A \in \cS(X\ox A).
\]

The entropy for quantum states $\rho\in \cS(A)$ is the von Neumann entropy, $S(\rho^A)=S(A)_\rho=-\Tr\rho\log\rho$, and by analogy with the Shannon entropy we have also the conditional entropy $S(A|B)_\rho = S(AB)_\rho-S(B)_\rho$ and the mutual information $I(A:B)_\rho = S(A)_\rho+S(B)_\rho-S(AB)_\rho$.

A special class of quantum channels are \emph{classical-quantum (cq-)channels}. These are given by $W:X\rightarrow B$, with $W(\ket{x}\!\bra{x'}) = \delta_{xx'}W_x$ for states $W_x\in\cS(B)$, $x\in\cX$. As the channel is entirely described by this family of states, we shall identify a cq-channel with the map $W:\cX\rightarrow\cS(B)$, mapping $x\mapsto W_x$. 

We shall need two commonly used measures of distance and similarity of quantum states. The fidelity of two quantum states $\rho$ and $\sigma$ is defined as 
$F(\rho,\sigma)=\|\sqrt{\rho}\sqrt{\sigma}\|_1 = \Tr\sqrt{\sqrt{\rho}\sigma\sqrt{\rho}}$, where $\|X\|_1 = \Tr\sqrt{XX^\dagger}$ is the trace norm. Then, according to Fuchs and van de Graff, 
\[
  1-F(\rho,\sigma) \leq \frac12\|\rho-\sigma\|_1 \leq \sqrt{1-F(\rho,\sigma)^2}.
\]

We end this section by recalling the formula for the private capacity of a quantum wiretap channel \cite{Ca/Wi/Ye,De} when no correlation is present, i.e. $\zeta^{A'B'E'} = \proj{0}^{A'} \ox \proj{0}^{E'} \ox \proj{0}^{E'} =: \emptyset$: 
\begin{equation}
P(\cN) = P(\cN,\emptyset)
       =\sup_n \frac1n \sup_{q(u),\rho_u} \left( I(U:B^n)_\gamma-I(U:E^n)_\gamma\right), 
\end{equation}
where the inner maximisation is over ensembles of states $\rho_u \in \cS(A^n)$ with probabilities $q(u)$, $u\in\cU$, and
\[
  \gamma^{UB^nE^n} = \sum_{u\n\cU} q(u) \proj{u}^U \ox \left[\cN^{\ox n}(\rho_u)\right]^{B^nE^n}.
\]

\section{Main result}
\label{sec:main}
Looking again at the definitions and the communication diagram in Fig. \ref{wireqresouima2}, we note that not just a code but any given modulations 
$\cE_u$ ($u\in\cU$) give rise to a cq-wiretap channel
$\cM:\cU \rightarrow (BB')^n(EE')^n$, mapping $u\in\cU$ to 
$\cM(u) = (\cN^{\ox n}\!\circ\cE_u)\zeta^{\ox n}$. We can thus apply the coding 
theorem and weak converse from \cite{Ca/Wi/Ye} to get the following 
expression for the assisted private capacity:
\begin{equation}
  \label{eq:private-trivial}
  P(\cN,\zeta) = \sup_n \frac1n \sup_{\{q(u),\cE_u\}} \!\!\left( I(U\!:\!B^n{B'}^n)_\gamma-I(U\!:\!E^n{E'}^n)_\gamma \right) \!, 
\end{equation}
with respect to the cq-state
\[
  \gamma^{U(BB')^n(EE')^n} = \sum_u q(u) \proj{u}^U \ox (\cN^{\ox n}\!\circ\cE_u)\zeta^{\ox n},
\]
where the inner supremum is over all alphabets $\cU$, probability distribution $q$ on $\cU$ and modulations $\cE_u$. Our goal will be to find a better achievability bound, which thus speeds up the convergence of Eq. \eqref{eq:private-trivial}. 

For the subsequent analysis, we observe that w.l.o.g. the reduced state $\zeta^{A'}$ has full rank (i.e. equal to the Hilbert space dimension $|A'|$), because otherwise we can shrink $A'$ to the support of $\zeta^{A'}$ and restrict the modulation maps $\cE_u$ accordingly. 

\begin{figure}[ht]
\begin{center}
\includegraphics[width=0.99\linewidth]{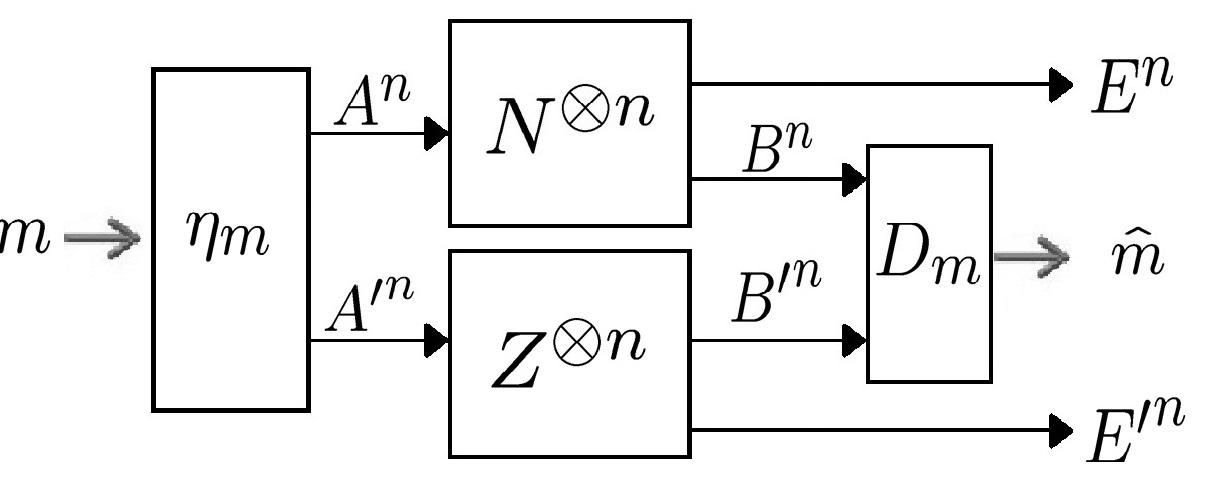}
\caption{Reformulation of the assisted wiretap code for $(\cN,\zeta)$ in terms of the tensor product wiretap channel $\cN\ox\cZ$ with a restriction on the marginals of the signal states $\eta_m$ on ${A'}^n$.}
\label{wireqres-alt}
\end{center}
\end{figure}

\begin{remark}
\label{rem:reformulation}
\normalfont
The key to our coding theorem is the following equivalent description of the code. Consider a Hilbert space $A''\simeq A'$ and a pure state vector $\ket{\phi_0}\in A'\ox A''$ such that $\phi_0^{A'} = \phi_0^{A''} = \zeta^{A'}$. By the generalised Choi theorem, there is a (unique) cptp map $\cZ:A'\rightarrow B'\ox E'$ such that 
$\zeta^{A'B'E'} = (\id_{A'}\ox\cZ^{A''\rightarrow B'E'})\phi_0$. 
Likewise, the generalised Choi states 
$\eta_u = (\cE_u\ox\id_{{A''}^n})\phi_0^{\ox n}$ of the modulation maps 
have the property 
$\Tr_{A^n}\eta_u = (\zeta^{A'})^{\ox n}$ and by these two equations 
determine $\cE_u$ uniquely as cptp maps. 

Now we make the elementary but crucial observation that
\begin{equation}\begin{split}
  \label{eq:eta-Z-swap}
  (\cE_u\ox\id_{{B'}^n{E'}^n})\zeta^{\ox n} &= (\id_{A^n}\ox\cZ^{\ox n})\eta_u, \\
  (\cN^{\ox n}\circ\cE_u\ox\id_{{B'}^n{E'}^n})\zeta^{\ox n}
                                            &= (\cN\ox\cZ)^{\ox n}\eta_u,
\end{split}\end{equation}
which means that our assisted wiretap code for $(\cN,\zeta)$ turns out to be equivalent to a regular wiretap code for $\cN\ox\cZ:AA' \rightarrow BB'\ox EE'$, 
albeit with the restriction $\Tr_{A^n}\eta_m = (\zeta^{A'})^{\ox n}$ for all
messages $m$ (see Fig. \ref{wireqres-alt}), 
and similarly for arbitrary modulations $\cE_u$.
\end{remark}

We can now relax the condition $\Tr_{A}\eta_u = \zeta^{A'}$ for all $u\in\cU$
to an average one, $\sum_u q(u) \Tr_A \eta_u = \zeta^{A'}$, to obtain the following 
coding theorem. This is where the Gel'fand-Pinsker and ``dirty paper'' coding idea is used \cite{Ge/Pi,Co}.

\begin{theorem}
\label{thm:wiretap-improved}
Let $(\cN,\zeta)$ be an assisted quantum wiretap channel with the wiretap channel 
$\cZ:A'\rightarrow B'E'$ defined as above. Assume furthermore a cq-channel 
$\cC:\cU \rightarrow A\ox A'$ and probabilities $q(u)$ such that 
$\sum_u q(u) \Tr_A\cC(u) = \zeta^{A'}$. Then,
\begin{equation}
  P(\cN,\zeta) \geq I(U:BB')_\gamma - \max\left( I(U:EE')_\gamma,I(U:A')_\beta \right), 
\end{equation}
where 
\begin{align}
  \label{eq:beta}
  \beta^{UAA'}     &\!= \sum_u q(u) \proj{u}^U \!\ox \cC(u)^{AA'}, \\
  \label{eq:gamma}
  \gamma^{UBB'EE'} &\!= \sum_u q(u) \proj{u}^U \!\ox \left((\cN\ox\cZ)\cC(u) \right)^{BB'EE'}\!.
\end{align}
\end{theorem}

Before proving this, let us remark that it includes the achievability part of
Eq. \eqref{eq:assisted-private}, since we may choose $\cC(u) = \eta_u = (\cE_u\ox\id_{A''})\phi_0$ as in Remark \ref{rem:reformulation}, which results in $I(U:A')_\beta=0$. 

\begin{proof}
We will construct $\eta_m$ on block length $n$ as in Remark \ref{rem:reformulation} by employing the wiretap coding strategy of Devetak \cite{De} for the cq-wiretap channel $\cM = (\cN\ox\cZ)\circ\cC : \cU \rightarrow BB'\ox EE'$, and the additional eavesdropper channel $\cU\ni u \mapsto \Tr_A\cC(u)$.
This amounts to choosing code words $u^n\in\cU^n$ independently according to the distribution $q^{\ox n}$, and binning them randomly so as to average over bins. 

Concretely, let $\cU^n \ni u_{(m,s)} \sim q^{\ox n}$ be sampled i.i.d. for 
$m\in[M]$ and $s\in[S]$, where $S = 2^{n\max(I(U:EE'),I(U:A'))+n\epsilon}$
and $MS = 2^{nI(U:BB')-n\epsilon}$ for an arbitrary $\epsilon>0$. (Devetak 
restricts the sampling to the typical sequences of $q^{\ox n}$, but by 
typicality the difference to the present prescription is small.) 

Devetak's analysis \cite{De} implies that for sufficiently large $n$, 
with high probability Bob's output 
states $\Tr_{E^n{E'}^n}\cM^{\ox n}(u_{(m,s)})$ are distinguishable reliably by a decoding POVM, say with error probability $\epsilon$. 
Furthermore, the averaged input states 
\[
  \widetilde{\eta}_m := \frac1S \sum_{s=1}^S \cC^n(u_{(m,s)}) \in \cS(A^n\ox{A'}^n)
\]
with high probability have the properties
\begin{align}
  \frac1M\! \sum_{m=1}^M \left\| \widetilde{\eta}_m^{{A'}^n} \!- (\zeta^{A'})^{\ox n} \right\|_1 &\leq \epsilon, \\
  \frac1M\! \sum_{m=1}^M \left\| (\Tr_{BB'}\circ(\cN\ox\cZ))^{\ox n}\widetilde{\eta}_m \!- (\gamma^{EE'})^{\ox n} \right\|_1 &\leq \epsilon,
\end{align}
where $\gamma$ is as in Eq. \eqref{eq:gamma}. The first comes from the 
Holevo-Schumacher-Westmoreland theorem for classical information transmission 
over a quantum channel, 
the second from the matrix tail bounds and the matrix covering lemma 
in \cite{Ahl/Win}.

By the Fuchs-van-de-Graaf relations between trace distance and fidelity, and
Uhlmann's theorem (cf. \cite{Ni/Ch,Wil}), we can find 
$\eta_m\in\cS(A^n\ox{A'}^n)$ such that $\eta_m^{{A'}^n} = (\zeta^{A'})^{\ox n}$ for all $m$ and 
\begin{equation}
  \label{eq:eta-put-right}
  \frac1M \sum_{m=1}^M \left\| \widetilde{\eta}_m - \eta_m \right\|_1 \leq 4\sqrt{\epsilon},
\end{equation}
and furthermore 
\begin{equation}
  \label{eq:eta-private}
  \frac1M\! \sum_{m=1}^M \left\| (\Tr_{BB'}\circ(\cN\ox\cZ))^{\ox n}\eta_m \!- (\gamma^{EE'})^{\ox n} \right\|_1 \leq \epsilon + 4\sqrt{\epsilon}.
\end{equation}

According to Remark \ref{rem:reformulation}, this is equivalent to an assisted 
wiretap code for $(\cN,\zeta)$ with block length $n$, $\lambda = \mu = \epsilon + 4\sqrt{\epsilon}$, concluding the proof.
\end{proof}

\section{Discussion}
\label{sec:conclusion}
We have presented a coding strategy for the quantum wiretap channel assisted by 
a general quantum correlation, which shows a systematic improvement of the 
private communication rate over the unassisted private capacity of the channel 
due to the exploitation of the correlation. 
Our Theorem \ref{thm:wiretap-improved} also directly generalises the main result of \cite{Ch/Ca/Se}, since classical channels are a special case of cptp maps and a classical correlation between random variables $X$, $Y$ and $Z$ is naturally represented by the diagonal state $\zeta^{A'B'E'} = \sum_{xyz} P_{XYZ}(xyz) \proj{x}^{A'} \ox \proj{y}^{B'} \ox \proj{z}^{E'}$.
As the proposed code uses a method 
derived from the (quantum) Gel'fand-Pinsker wiretap channel with side information, 
it is also always at least as good as, or superior to, the ``trivial'' 
incorporation of the correlation resource into an augmented wiretap channel. 
However, due to the general need for regularisation, both our main result 
(Theorem \ref{thm:wiretap-improved}) and the ``trivial'' achievable rate 
[Eq. \eqref{eq:private-trivial}] lead to multi-letter formulas for the same 
quantity $P(\cN,\zeta)$.

Of course, there are extreme cases of useless correlation: for instance any 
state $\zeta$ such that Alice and Bob are uncorrelated, 
$\zeta^{A'B'} = \zeta^{A'}\ox\zeta^{B'}$, clearly leads to
$P(\cN,\zeta)=P(\cN)$, irrespective of the wiretap channel $\cN$.
On the other hand, any $\zeta^{A'B'}\ox\proj{0}^{E'}$ that is not a tensor product 
between Alice and Bob offers an advantage for some channel, indeed we can take the 
quantum broadcast channel $\cB:A\rightarrow B\ox E$ with qubits $A$, $B$ and $E$, 
acting as $\cB(\rho) = B\rho B^\dagger$, $B\ket{x}=\ket{x}^B\ket{x}^E$ for 
$x=0,1$. Because $\cB$ is both degradable and anti-degradable, $P(\cB)=0$. 
However, $P(\cB,\zeta) > 0$, since $\cB$ can be used to communicate classically 
and thus Alice and Bob can extract shared randomness that is automatically
a secret key from $\zeta^{A'B'}$ \cite{De/Wi}, 
which then can be used to one-time-pad a subsequent classical message. 

As indicated in the introduction, various communication problems considered 
in the past are special cases of our model, and it is interesting to see 
to which extent our main theorem reproduces known results or sheds new light 
on them. 
For instance, the channel without wiretapper, 
$\cN^{A\rightarrow BE} = \cN^{A\rightarrow B}\ox\proj{0}^E$, gives rise to all 
sorts of problems of classical communication assisted by entanglement. 
For general point-to-point channel $\cN:A\rightarrow B$ but arbitrary pure 
state $\zeta^{A'B'}\ox\proj{0}^{E'}$, we recover the entanglement-assisted 
classical capacity $C_E(\cN)$ \cite{Be/Sh/Sm/Th,Be/Sh/Sm/Th2}, even though 
some additional work is still required to see how Theorem \ref{thm:wiretap-improved}
and Eq. \eqref{eq:assisted-private} give rise to the familiar maximum 
quantum mutual information formula from those papers. 

If instead our correlation resource is a mixed state, albeit tensor product 
with Eve, i.e. $\zeta^{A'B'}\ox\proj{0}^{E'}$, we get the classical capacity 
$C(\cN,\zeta)$ of $\cN:A\rightarrow B$ assisted by $\zeta^{A'B'}$ 
considered in \cite{Ba/Wi/Ya}. In that paper it was observed that a 
separable $\zeta^{A'B'}$ cannot increase the classical capacity of any channel,
$C(\cN,\zeta)=C(\cN)$, and the hypothesis was advanced that for every 
entangled $\zeta$ there might be a channel such that $C(\cN,\zeta)>C(\cN)$. 
Noticing that in our present setting, and due to the triviality of Eve in 
both the channel and the resource state, the classical capacities (assisted 
and unassisted) are equal to private capacities, suggests a broader interpretation 
of the question from \cite{Ba/Wi/Ya}: namely, for any state 
$\rho^{A'B'}$ and a purification $\zeta^{A'B'E'}$ of it, is it true 
that $\rho^{A'B'}$ is entangled if and only if 
there exists a wiretap channel $\cN:A\rightarrow BE$ such that 
$P(\cN,\zeta)>P(\cN)$? 

The special case of the ideal channel $\cN=\id_A:A\rightarrow B=A$ merits 
a separate mention, too. Its assistance by a general product state 
$\zeta^{A'B'}\ox\proj{0}^{E'}$ was considered in multiple works
\cite{Bo/Pl/Ve,Bow,Hiro,Ho/Ho/Ho/Le/Te,Win2001,Ho/Pi}, and it was found that 
$C(\id_A,\zeta) \geq \log|A| + I(A\rangle B')_\omega$, where 
$\omega^{AB'} = (\Omega\ox\id_{B'})\zeta$ with a cptp map $\Omega:A'\rightarrow A$.
The optimisation of the coherent information $I(A\rangle B')_\omega$ over all 
systems $A$ and channels $\Omega$ leads to the \emph{dense coding advantage} $\Delta(A'\rangle B')_\zeta$ of the resource state: 
\[
  \Delta(A'\rangle B')_\zeta 
    := \max_{\Omega \text{ cptp}} 
      I(A\rangle B')_\omega 
      \text{ s.t. } 
      \omega^{AB'} = (\Omega\ox\id_{B'})\zeta.
\]
For sufficiently large $A$ system, it follows that 
$C(\id_A,\zeta) = \log|A| + \Delta^{(\infty)}(A'\rangle B')_\zeta$, where 
$\Delta^{(\infty)}$ is the regularised dense coding advantage, 
cf. \cite{Ho/Pi}:
\[
  \Delta^{(\infty)}(A'\rangle B')_\zeta 
    = \sup_n \frac1n \Delta({A'}^n\rangle {B'}^n)_{\zeta^{\otimes n}}.
\]
Now this achievability bound and expression for the 
assisted capacity correspond exactly to Eq. \eqref{eq:assisted-private}, 
and interestingly it provides an instance where Theorem \ref{thm:wiretap-improved}
is better, at least on the single-letter level (naturally, in the regularisation both expression result in the same number). 
Namely, in \cite{Ho/Pi} a certain duality relation was observed between the dense coding advantage 
and the so-called \emph{entanglement of purification} $E_P$ \cite{EoP}, which extends to the 
regularisations of the two quantities: 
\[
  E_P(C:D)_\rho := \min_{\cT:E\rightarrow F} S(BF)_\omega,
\]
for a bipartite state $\rho^{CD}$ with purification $\psi^{CDE}$, where the minimisation is over cptp maps $\cT:E\rightarrow F$ and $\omega^{BF}=(\id_B\ox \cT)\psi^{BE}$. 
Namely, for any purification $\psi^{A'B'C'}$ of $\zeta^{A'B'}$, 
\[
  \Delta(A'\rangle B') + E_P(C':B') = S(B').
\]
It remains unknown whether  $\Delta=\Delta^{(\infty)}$ and $E_P=E_P^{(\infty)}$, but in \cite{Che/Wi} numerical evidence to the contrary was provided for the entanglement of purification. 
Showing $E_P(\rho) > E_P^{(\infty)}(\rho)$
for a certain state in \cite{Che/Wi} required calculating the l.h.s. numerically,
and finding a new upper bound for the r.h.s. via an asymptotic protocol using the covering lemma. Concretely, it was shown for an ensemble $\{q(u),\rho_u^{CD}\}$ with $\rho=\sum_u q(u)\rho_u$ that 
\[
  E_P^{(\infty)}(C:D)_\rho 
    \leq \sum_u q(u) E_P(C:D)_{\rho_u} + I(U:CD)_\gamma,
\]
where $\gamma^{UCD} = \sum_u q(u)\proj{u}^U \ox \rho_u^{CD}$ is the cq-state of the ensemble.
Using the duality relation from \cite{Ho/Pi} to translate this protocol to $\Delta^{(\infty)}$, it corresponds to the achievable rate from Theorem \ref{thm:wiretap-improved} in the case $\cN=\id_A$ and $\zeta^{A'B'}$.

\section*{Acknowledgments}
MLC was supported  by the German Research
Foundation (DFG) under the Walter Benjamin Fellowship CA-2779/1-1, 
and in part by the ESA SATNEX V programme (project 4000130962/20/NL/NL/FE).
Both authors acknowledge support by the Spanish MICIN (project PID2022-141283NB-I00) with the support of FEDER funds.
AW is furthermore supported by the European Commission QuantERA grant ExTRaQT (Spanish MICIN project PCI2022-132965), the Spanish MICIN with funding from European Union NextGenerationEU (PRTR-C17.I1) and the Generalitat de Catalunya, by the Alexander von Humboldt Foundation, and the Institute for Advanced Study of the Technical University Munich.

\end{document}